\documentclass[fleqn,usenatbib]{mnras}

\usepackage{newtxtext,newtxmath}
\usepackage[T1]{fontenc}
\usepackage{ae,aecompl}

\usepackage{graphicx}	% Including figure files
\usepackage{amsmath}	% Advanced maths commands
\usepackage{amssymb}	% Extra maths symbols
\usepackage{tabularx}
\newcommand{\Gaia}{\textit{Gaia}}
\newcommand{\teff}{$\mathrm{T_{eff}}$}
\newcommand{\logg}{$\log g$}
\newcommand{\Nstar}{NGTS J2143-38 }
\newcommand{\Nstarlong}{NGTS J214358.5-380102 }
\newcommand{\Nstarlongns}{NGTS J214358.5-380102}
\newcommand{\Nstarns}{NGTS J2143-38}
\newcommand{\NstarA}{NGTS J2143-38A }
\newcommand{\NstarB}{NGTS J2143-38B }

\newcommand{\NstarRadA}{$0.461 ^{+0.038}_{-0.025}$} %Primary radius
\newcommand{\NstarRadB}{$0.411 ^{+0.027}_{-0.039}$} %Secondary radius
\newcommand{\NstarMassA}{$0.426 ^{+0.0056}_{-0.0049}$} %Primary mass
\newcommand{\NstarMassB}{$0.455 ^{+0.0058}_{-0.0052}$} %Secondary mass
\newcommand{\TeffA}{$3452.97 ^{+56.34}_{-50.34}$} %Primary Teff
\newcommand{\TeffB}{$3280.77 ^{+75.20}_{-126.74}$} %Secondary Teff

\newcommand{\eccerror}{$0.323^{+0.0014}_{-0.0037}$}
\newcommand{\period}{7.618}
\newcommand{\ecc}{0.323}
\newcommand{\incl}{87.587}
\newcommand{\impactparam}{1.431}
\newcommand{\pericentre}{10.615}

\newcommand{\masy}{mas\,yr$^{-1}$}
\newcommand{\ms}{ms$^{-1}$ }

\title[Most Eccentric Known M-Dwarf Binary]{\Nstarlong - NGTS discovery of the most eccentric known M-Dwarf binary system}

\author[J. S. Acton et al.]{
\parbox{\textwidth}{
Jack S.~Acton,$^{1}$\thanks{E-mail:ja466@le.ac.uk}
Michael R.~Goad,$^{1}$
Liam Raynard,$^{1}$
Sarah L.~Casewell,$^{1}$
James A. G.~Jackman,$^{2,3}$
Richard D.~Alexander,$^{1}$
David R.~Anderson,$^{2,3}$
Daniel Bayliss,$^{2,3}$
Edward M.~Bryant,$^{2,3}$
Matthew R.~Burleigh,$^{1}$
Claudia Belardi,$^{1}$
Benjamin F. Cooke,$^{2,3}$
Philipp Eigm\"uller,$^{10}$
Samuel Gill,$^{2,3}$
James S. Jenkins,$^{4,5}$
Monika Lendl,$^{6,7}$
Tom Louden,$^{2,3}$
James McCormac,$^{2,3}$
Maximiliano Moyano,$^{9}$
Louise D.~Nielsen,$^{6}$
Rosanna H.~Tilbrook,$^{1}$
St\'{e}phane~Udry,$^{6}$
Christopher A.~Watson,$^{8}$
Richard G. West,$^{2,3}$ 
Peter J.\ Wheatley$^{2,3}$
Jose I. Vines,$^{4}$
}
\\
$^{1}$School of Physics and Astronomy, University of Leicester, University Road, Leicester, LE1 7RH, UK\\
$^{2}$Dept. of Physics, University of Warwick, Gibbet Hill Road, Coventry CV4 7AL, UK\\
$^{3}$Centre for Exoplanets and Habitability, University of Warwick, Gibbet Hill Road, Coventry CV4 7AL, UK\\
$^4$Departamento de Astronom\'ia, Universidad de Chile, Camino el Observatorio 1515, Las Condes, Santiago, Chile\\
$^5$Centro de Astrof\'isica y Tecnolog\'ias Afines (CATA), Casilla 36-D, Santiago, Chile\\
$^{6}$Observatoire de Gen{\`e}ve, Universit{\'e} de Gen{\`e}ve, 51 Ch. des Maillettes, 1290 Sauverny, Switzerland\\
$^{7}$Space Research Institute, Austrian Academy of Sciences, Schmiedlstr. 6, 8042 Graz, Austria\\
$^{8}$Astrophysics Research Centre, School of Mathematics and Physics, Queen's University Belfast, BT7 1NN, Belfast, UK\\
$^{9}$Instituto de Astronom\'ia, Universidad Cat\'olica del Norte, Angamos 0610, 1270709 Antofagasta, Chile\\
$^{10}$Institute of Planetary Research, German Aerospace Center, Rutherfordstrasse 2, 12489 Berlin, Germany\\}

\date{Accepted XXX. Received YYY; in original form ZZZ}

\pubyear{2020}

\begin{document}
\label{firstpage}
\pagerange{\pageref{firstpage}--\pageref{lastpage}}
\maketitle

\begin{abstract}
We present the discovery of \Nstarlongns, an eccentric M-dwarf binary discovered by the Next Generation Transit Survey. The system period of \period ~days is greater than many known eclipsing M-dwarf binary systems. Its orbital eccentricity of \eccerror, is large relative to the period and semi-major axis of the binary. Global modelling of photometry and radial velocities indicate stellar masses of M$_{\rm A}$=\NstarMassA M$_{\odot}$, M$_{\rm B}$=\NstarMassB M$_{\odot}$ and stellar radii  R$_{\rm A}$=\NstarRadA R$_{\odot}$, R$_{\rm B}$=\NstarRadB R$_{\odot}$, respectively. Comparisons with stellar models for low mass stars show that one star is consistent with model predictions whereas the other is substantially oversized.  Spectral analysis of the system suggests a primary of spectral type M3V, consistent with both modelled masses and radii, and with SED fitting of NGTS photometry. As the most eccentric M-dwarf binary known, \Nstarlong provides an interesting insight into the strength of tidal effects in the circularisation of stellar orbits.
\end{abstract}


%%%%%%%%%%%%%%%%%%%%%%%%%%%%%%%%%%%%%%%%%%%%%%%%%%

%%%%%%%%%%%%%%%%% BODY OF PAPER %%%%%%%%%%%%%%%%%%

\section{Introduction}
M-dwarfs are the most common stars in the galaxy (\citealt{Henry2006, Bochanski2010}) and provide a promising environment in which to search for Earth or super-Earth sized planets. Recent investigations of nearby M-dwarfs have discovered that some of these stars host multiple planet systems (\citealt{Gillon2017, Guenther2019, Kostov2019}). Since all of these systems are nearby, within  $\sim 20$~pc, this would suggest that compact multi-planet systems around M-dwarf stars are likely common.

Characterisation of newly discovered exoplanets is important for determining the area of parameter space that they occupy among the $\approx$~4000 (as of November 2019; NASA exoplanet archive) currently known exoplanets. However, in order to fully understand any discovered planetary system, we must first understand the host star. In particular, accurate knowledge of the mass and radius is vital for the accurate determination of the mass and radius (and hence the bulk density) of any transiting exoplanet. 

Unfortunately, M-dwarfs are much less well understood than the F-, G- and K-type stars that are the focus of many exoplanet surveys (e.g., Kepler \citealt{Basri2005}, WASP \citealt{Pollacco2006}). For example, it has been shown that measured masses and radii for M-dwarfs may differ from model predictions by up to 10 per cent (\citealt{Feiden2012, Terrien2012}). In exoplanet characterisation these models are used to determine the mass of the host star, and thus infer the structure of the planet. If the models are inaccurate, then this will impact on the derived parameters for any planet discovered around a low mass star. Measurements from the $Gaia$ mission (\citealt{GAIA2016}) mean that many stars now have measured radii derived from their luminosity and spectral type (\citealt{Morrell2019}). However, M-dwarf masses can only be directly measured when they occur in binary systems.  In this regard, eclipsing binaries are especially important, as they yield directly both mass and radius estimates. For the remainder, secondary mass indicators (e.g., from $\log g$ and spectral typing) must be used.

This is further complicated by the fact that the deviation between models and measurements is not consistent between systems. \cite{Boyajian2012} show that the discrepancy between mass-radius measurements and model predictions of eclipsing binary systems may be dependent on the period of the binary, with the largest discrepancies (in mass and radius) found for short period systems. Yet, many of the known eclipsing M-dwarf binaries are in short period orbits of little more than a day (e.g., \citealt{Casewell2018, Morales2009}), while the parameter space for longer period binaries is much more sparse. \cite{Parsons2018} list 33 well characterised double lined M/M eclipsing binaries, only 2 of which have orbital periods greater than 5 days; the vast majority having periods less than 2 days. Most M-dwarfs in longer period orbits are in systems consisting of an M-dwarf secondary with a more massive stellar companion(e.g., \citealt{Gill2019, Lendl2019}).

It was shown by \cite{Torres2013} that M-Dwarfs in detached eclipsing binaries tend to also have cooler than expected effective temperatures. Cooler temperatures,  when combined with the aforementioned inflated radii of M-dwarfs can result in stellar luminosities that are consistent with models. This may suggest that effects not accounted for in models arise due to effects of the surface of the stars themselves, such as metallicity (\citealt{Stassun2012},\citealt{Lopez2007}) or stellar activity \citep{Stelzer2013}. The latter effect may be more prevalent in short period systems, which are tidally locked, rapidly rotating and magnetically active, resulting in the star being larger and cooler than predicted by models (\citealt{Morales2008}). This hypothesis is supported by the work of \citealt{Demory2009}, who showed that interferometric radii of isolated low mass stars are consistent with model predictions.

Curiously, the discrepancy between mass-radius measurements and model predictions is not an effect caused purely by the binarity of the stars. Both models (e.g \citealt{Spada2013}) and observations (e.g \citealt{Boyajian2012}) have shown that this discrepancy is present in both isolated and binary M-Dwarfs. The suggestion that the discrepancy in values derived from eclipsing binaries is caused by the effect of short orbital periods is also disputed. There are examples of longer period systems, which we would not expect to be tidally locked, that show the same over sizing as these short period systems (\citealt{Doyle2011}, \citealt{Irwin2011}), as well as short period systems that show good agreement with models (\citealt{Blake2008}). 

It should also be noted that whilst the majority of known M-dwarf binaries are short period systems in tight circular orbits, many binary stars are located in non-circular orbits \citep{Bulut2007}. This can contribute to increased activity on the stellar surfaces at and around periastron phase \citep{Moreno2011}. This increased and variable activity may have some effect on the size of the stars in the binary, contributing to a discrepancy between models and measured parameters. However there are very few known M-Dwarfs in eccentric orbits, \cite{Parsons2018} list only 15 with $e>0.1$, only three of which are double M-Dwarf systems. With so few of these types of systems known, it is difficult to quantify the significance of this effect, if any.

It is clear then that we need to understand and characterise a wide variety of M-dwarf binary systems, in order to be able to properly constrain the mass/radius relation for these low mass stars. To do so we need to compile a large sample of M-dwarf eclipsing binaries for study. The most recent exoplanet surveys provide an ideal opportunity to do this, as the photometric signal produced by an eclipsing binary is similar in nature to that of a transiting exoplanet. Thus these surveys are able to detect large numbers of these systems suitable for further study.

\subsection{Low Mass stars in NGTS}

The Next Generation Transit Survey (hereafter NGTS; \citealt{Wheatley2018}) is particularly well-suited to the study of low mass stars. Comprising an array of 12 fully automated 20cm telescopes operating at ESO's Paranal observatory in Chile \citep{Wheatley2018}, this wide field (instantaneously covering 96 sq deg) red-sensitive survey, routinely delivers high cadence (every 12 seconds), high precision ($\sim$ mmag) photometry for tens of thousands of stars, including numerous examples of early--mid M dwarfs. NGTS has been operational since early 2016 and is optimized for detecting small planets around K- and M-type stars (e.g \citealt{Bayliss2018},\citealt{West2019}). It is expected that NGTS will discover $\sim 300$ new exoplanets and $\sim 5600$ eclipsing binaries during the mission lifetime \citep{Guenther2017a}.

 The NGTS filter was specifically designed to be sensitive at the red end of the spectrum, with a bandpass range of 520 to 890 nm. This particular wavelength range allows for good red sensitivity whilst also avoiding strong water absorption bands present beyond 900~nm. Additionally, the design of NGTS allows for increased sensitivity (mmag precision) for stars fainter than $m_{I}\approx12.5$ which is of great value for M-dwarf stars which tend to be relatively faint. NGTS has already discovered the most massive planet orbiting an M-type star, NGTS-1b \citep{Bayliss2018}, as well as the the shortest period transiting brown dwarf orbiting a main sequence star, an active M-Dwarf \citep{Jackman2019}. These stars continue to be a key focus of our planet search.

Of particular interest are those M-dwarf stars that reside in eclipsing binaries, as they allow accurate masses to be determined via radial velocity (RV) measurements. This, when combined with measurements of stellar radii from eclipse light curves allows for accurate characterisation of the component stars. NGTS has already identified one of only three  field-age ($<$1 Gyr) late M-dwarf binaries \citep{Casewell2018} which provides a valuable addition to the mass-radius relationship for M-dwarfs. NGTS has also identified many candidate eclipsing binary systems covering a wide range of parameter space which continue to be investigated (e.g \citealt{Gill2019}, \citealt{Lendl2019}).

In this paper we present the NGTS discovery of one such interesting M-dwarf binary, \Nstarlong (hereafter \Nstar). We make use of follow-up photometry and radial velocity measurements to derive accurate orbital parameters for the system. We also use low resolution spectroscopy in order to accurately determine the spectral type of the system and allow for greater understanding of the context in which this discovery sits.

\begin{table}
	\centering
	\caption{Stellar Properties for \Nstar}
	\label{tab:stellar}
	\begin{tabular}{ccc} % six columns, alignment for each
	Property	&	Value		&Source\\
	\hline
	Gaia I.D.		&	DR2 6586032117320121856	& Gaia \\
    R.A.		&	21:43:58.6 				&	NGTS	\\
	Dec			&	-38:01:02.65	&		  NGTS \\
    $\mu_{\alpha}$ (\masy)& $26.186 \pm0.089$  & Gaia\\
    $\mu_{\delta}$ (\masy) & $-41.862 \pm0.093$ & Gaia\\
    dist(pc)&	$120.31 \pm1.02$& Gaia\\
    $G$			&$14.270872$	&Gaia\\
    NGTS		&$13.59$				& NGTS\\
    $J$			&$11.943\pm0.025$		&2MASS	\\
   	$H$			&$11.315\pm0.026$		&2MASS	\\
	$K_{s}$			&$11.046\pm0.023$		&2MASS	\\
    
	\hline
	\end{tabular}
\end{table}

\section{Observations}

\begin{figure}
	\includegraphics[width=\columnwidth]{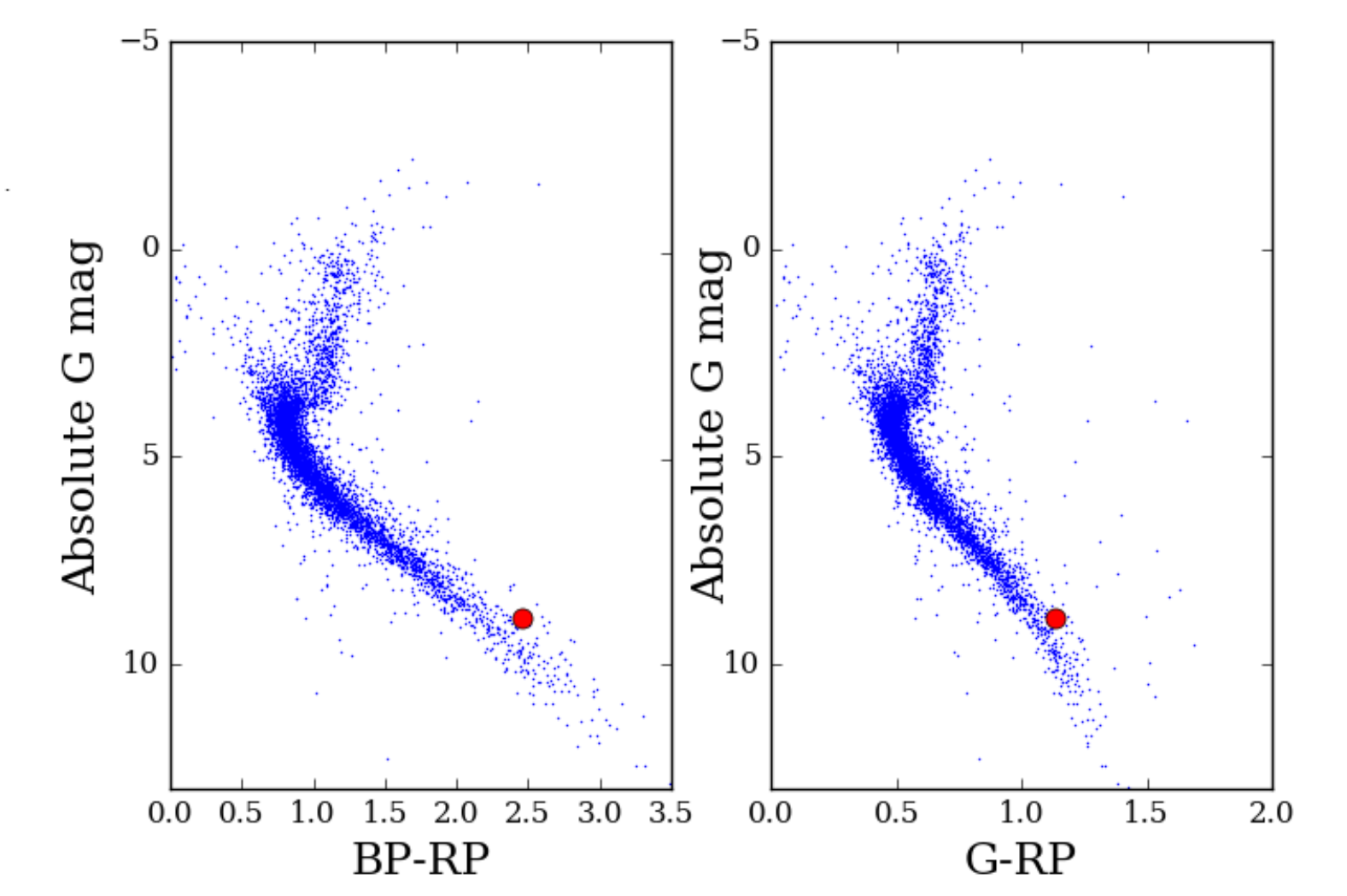}
    \caption{Hertzsprung-Russell diagram of stars in the NGTS field showing the position of \Nstar above the main sequence. This suggests it is likely a binary, as its position on the diagram makes it $\sim$ factor 2 more luminous than expected for a single star of similar spectral type.}
   \label{fig:HR_diagram}
\end{figure}
\begin{table*}
	\centering
	\caption{{Summary of observations.}}
	\label{tab:obs_summary}
	\begin{tabular}{ccccccc} % six columns, alignment for each
    \hline
Observation type & Telescope & Band  & Cadence & Total integration time & Period & Notes\\
\hline
Photometry	& NGTS	&	520-890\,nm	&	12\,s & 150\,nights	&	21/04/16-22/12/16 & 3 eclipses in total\\
Photometry	& SAAO	&	z'	&	60\,s & 5 hours	&	16/07/19 & 1 eclipse observed\\
Photometry	& TESS	&	600-1000\,nm	&	120\,s & 28\,nights	&	25/07/18-22/08/18 & 3 eclipses in total\\
Spectroscopy	& HARPS	&	378-691\,nm	& 1 hour &	1 hour	&	
24/07/17 & One RV Point \\
Spectroscopy	& CORALIE	&	390-680\,nm 	& 45mins-1 hour &	4.4 Hours	&	
29/08/17-16/09/19 & 5 RV Points  \\
Spectroscopy	& X-SHOOTER	&	NIR and VIS	& 120\,s &	960\,s	&	
02/09/18 & \\
%Spectroscopy	& ESO 3.6\,m (HARPS)	&	378-691\,nm & 1\,hour & %6\,hours	&	 & EGGS mode\\
	\hline
    \end{tabular}
\end{table*}

\Nstar was initially discovered as a periodic source in NGTS photometric lightcurves following the usual Box Least Squares (BLS; \citealt{Kovacs2016}) search for candidate transit events \citep{Wheatley2018}. Follow-up observations were performed with Sutherland High Speed Optical Cameras (SHOC) \citep{Coppejans2013} on the South African Astronomical Observatory (SAAO) 1-m telescope. This photometry was then used in conjunction with observations from the Transiting Exoplanet Survey Satellite (TESS) \citep{Ricker2014} for global modelling of the system. We obtained near-IR and visible wavelength spectra from X-SHOOTER [\citealt{Vernet2011},Very Large Telescope (VLT)] and radial velocity measurements from CORALIE [Swiss 1.2m Euler Telescope, \citealt{Queloz2000}] and HARPS [ESO 3.6m, \citealt{Mayor2003}]. These observations are detailed in Table \ref{tab:obs_summary} and described below.

\subsection{NGTS Photometry}
\begin{figure}
	\includegraphics[width=\columnwidth]{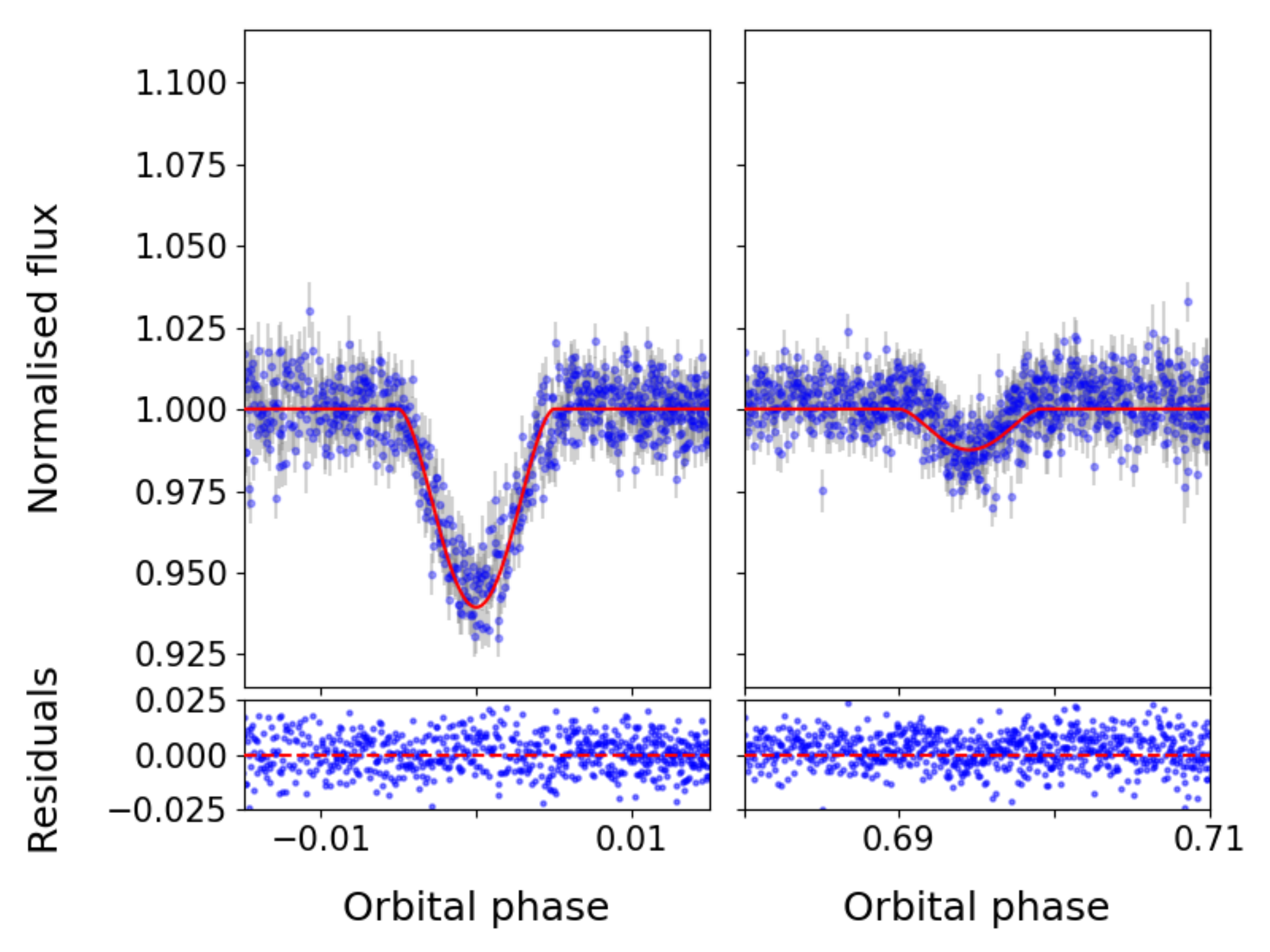}
    \caption{NGTS photometry of the primary and secondary eclipses of \Nstar folded on a period of \period\, days. The red line shows the model fit obtained from joint modelling of photometric and spectroscopic data.}
   \label{fig:ngts_phot}
\end{figure}
\begin{figure}
	\includegraphics[width=\columnwidth]{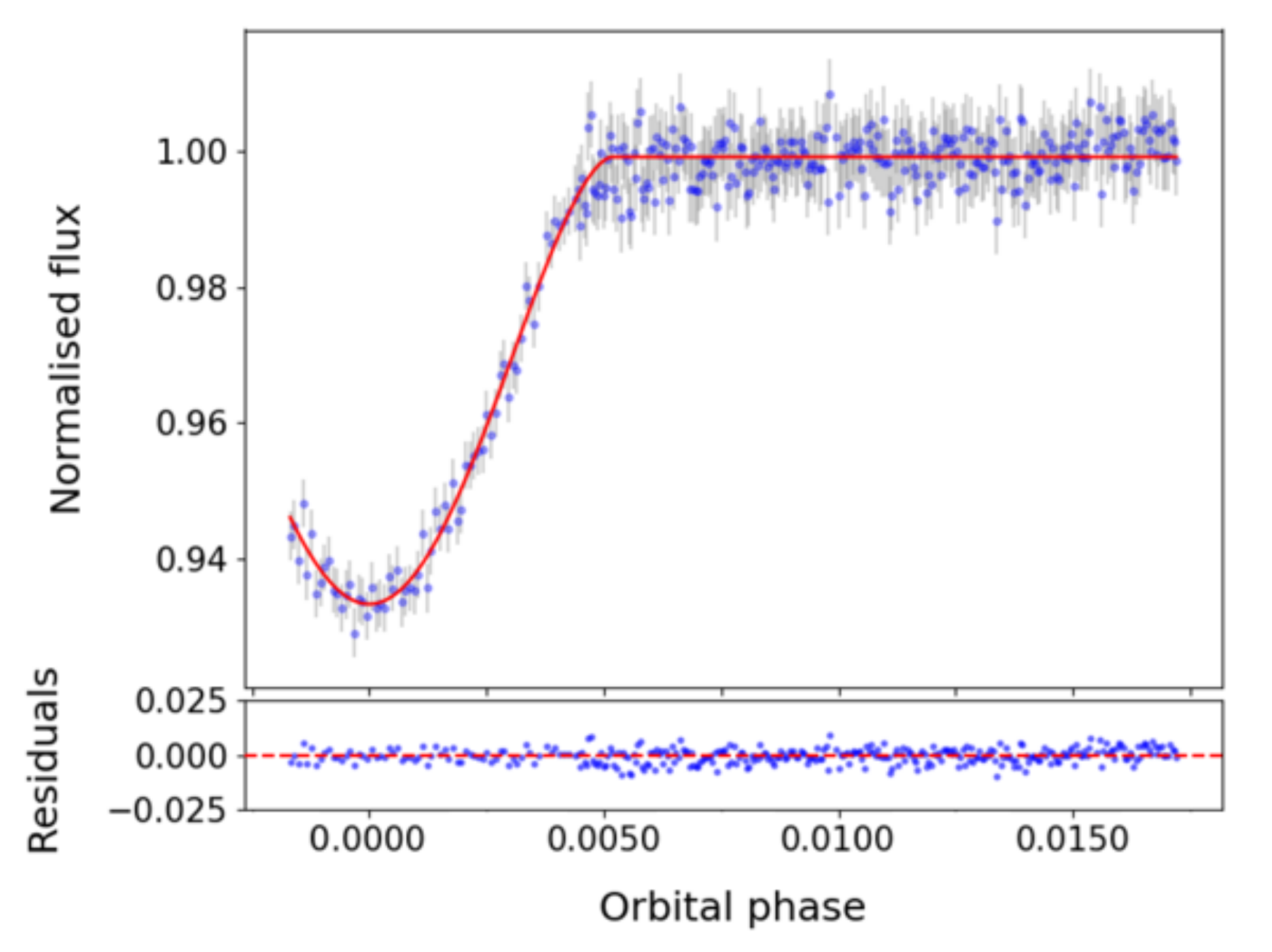}
    \caption{SAAO photometry of primary eclipse of \Nstar plotted in phase. Observations were taken on the 1m telescope using the $z'$ filter. The red line shows the model fit obtained from joint modelling of photometric and spectroscopic data.}
   \label{fig:saao_phot}
\end{figure}

The observations of \Nstar were taken during a standard NGTS survey season, spanning 150 nights between 21st April 2016 and 22nd December 2016. These observations cover 3 full primary eclipses, two partial eclipses and 4 secondary eclipses. The data for the entire field was reduced using the standard NGTS pipeline to produce detrended light curves suitable for classification.

The system was detected using \textsc{orion} \citep{Wheatley2018}, an implementation of the BLS algorithm \citep{Kovacs2016}, which provided some initial parameters for the system. \textsc{orion} identified a period of 7.617847 days, a transit width of 2.16~h and a depth of 3.3 per cent. It should be noted that this algorithm is optimized for planetary transit signals, which are usually more 'U'-shaped when compared to the 'V'-shaped eclipsing binary light curves. Consequently, it is possible that there is some discrepancy between parameter values reported by {\sc orion} and the true parameters of the system. Indeed upon inspection of the light curve we find that the true depth is around 6 per cent.

The object was identified based on its colour and proper motion as being a potentially interesting M-dwarf system. The phase folded light curve showed no secondary eclipse at phase 0.5 as would be expected for a stellar companion. The object was put forward for follow-up photometry and radial velocity measurements, which later confirmed it to be an eclipsing binary system, a common false positive signal found in exoplanet surveys. The binary nature of the system was further confirmed by its position above the main sequence on the Hertzsprung Russell diagram (see Figure~\ref{fig:HR_diagram}). At $\approx$ 0.75 magnitudes above the main sequence it is a factor 2 more luminous than a single star of the same spectral type (and indicative of a binary system comprising two roughly equal mass stars).

The primary and secondary eclipses in the NGTS lightcurve are shown in Figure~\ref{fig:ngts_phot} folded on the system period derived from global modelling. Note the secondary eclipse is positioned close to phase 0.7, rather than the expected position of phase 0.5. Assuming that the light curve has been folded on the correct period, this shows that the system is in an eccentric orbit.

\subsection{SAAO photometry}

\Nstar was observed by the SAAO 1-m telescope on 2017 July 17th. These observations were performed as part of the standard NGTS follow up program. The aim of the observations was to confirm any transit and more accurately determine the transit depth and width. Observations were conducted over $\sim 5$hrs using the SHOC camera \citep{Coppejans2013}, in the $z'$ band. 

The data were bias and flat field corrected via the standard procedure, using the \textsc{SAFPhot} Python package (Chaushev \& Raynard, in preparation). Differential photometry was also carried out using \textsc{SAFPhot}, by first extracting aperture photometry for both the target and comparison stars using the 'SEP' package \citep{Barbary2016}. The sky background was measured and subtracted using the SEP background map, adopting box size and filter width parameters which minimised background residuals, measured across the frame after masking the stars. A 32 pixel box size and 2 pixel box filter were found to give the best results. Three comparison stars were then utilised to perform differential photometry on the target, using a 6 pixel radius aperture which maximised the signal-to-noise.

SAAO observations captured the bottom of the eclipse, as well as the egress and some out-of-transit flux of the primary star. They indicate a depth that is consistent with NGTS photometry, and that the eclipse occurred at a time consistent with the ephemeris from \textsc{orion}. The reduced light curve is shown in Figure~\ref{fig:saao_phot}.

\subsection{TESS photometry}

\Nstar was observed by the Transiting Exoplanet Survey Satellite (hereafter TESS, \citealt{Ricker2014}) during Sector 1 of the mission. These observations occurred between 2019 July 25th and 2019 August 22nd. TESS observations typically consist of full frame images taken with 30 minute cadence, however 15,900 stars in Sector 1 were also observed at 2 minute cadence by Camera 1, including \Nstar (TIC ID = 197570458). The phase folded TESS light curve had a best BLS period of 7.618725 days, consistent with the period determined from NGTS observations.

\begin{figure}
	\includegraphics[width=\columnwidth]{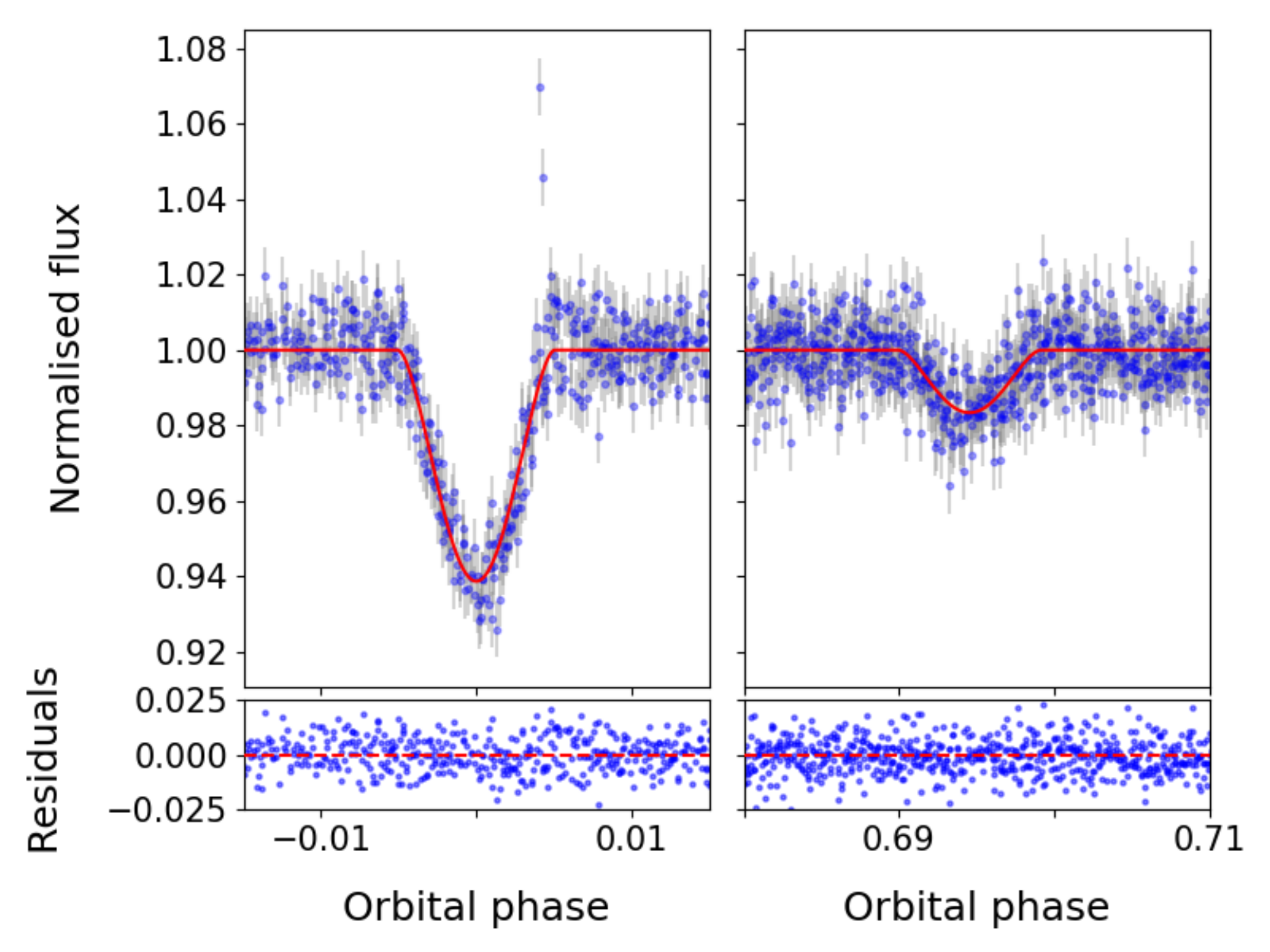}
    \caption{TESS photometry of primary and secondary eclipse of \Nstar folded on the period of \period\, days. The red line shows the model fit obtained from joint modelling of photometric and spectroscopic data.}
   \label{fig:tess_phot}
\end{figure}

TESS captured three primary and two secondary eclipses of the system during this observation window. The phase folded TESS photometry also shows a secondary eclipse feature at phase 0.7, as in the NGTS photometry. The detection of this feature by two independent surveys confirms that the binary is in an eccentric orbit. The primary and secondary eclipses of the TESS light curve, folded on the period derived from global modelling (see \S3.2), are shown in Figure~\ref{fig:tess_phot}.

\subsection{HARPS \& CORALIE Spectra}

\Nstar was observed with the CORALIE spectrograph \citep{Queloz2000} on the Swiss 1.2-metre Leonhard Euler Telescope at ESO's La Silla Observatory, Chile. The object is very faint for CORALIE (V=15.39), so we took long ($\sim 60$ min) exposures to maximise the signal-to-noise ratio.  This allowed for eventual precision in the radial velocity measurement of around 250 \ms which is adequate for analysis of this system due to the high radial velocity of the binary. We additionally obtained a single spectrum from HARPS \citep{Mayor2003}. We assume any systematic offset between the radial velocities for our instruments to be negligible given the magnitude of the radial velocity shift that is expected from a binary of this type.

\begin{figure}
	\includegraphics[width=\columnwidth]{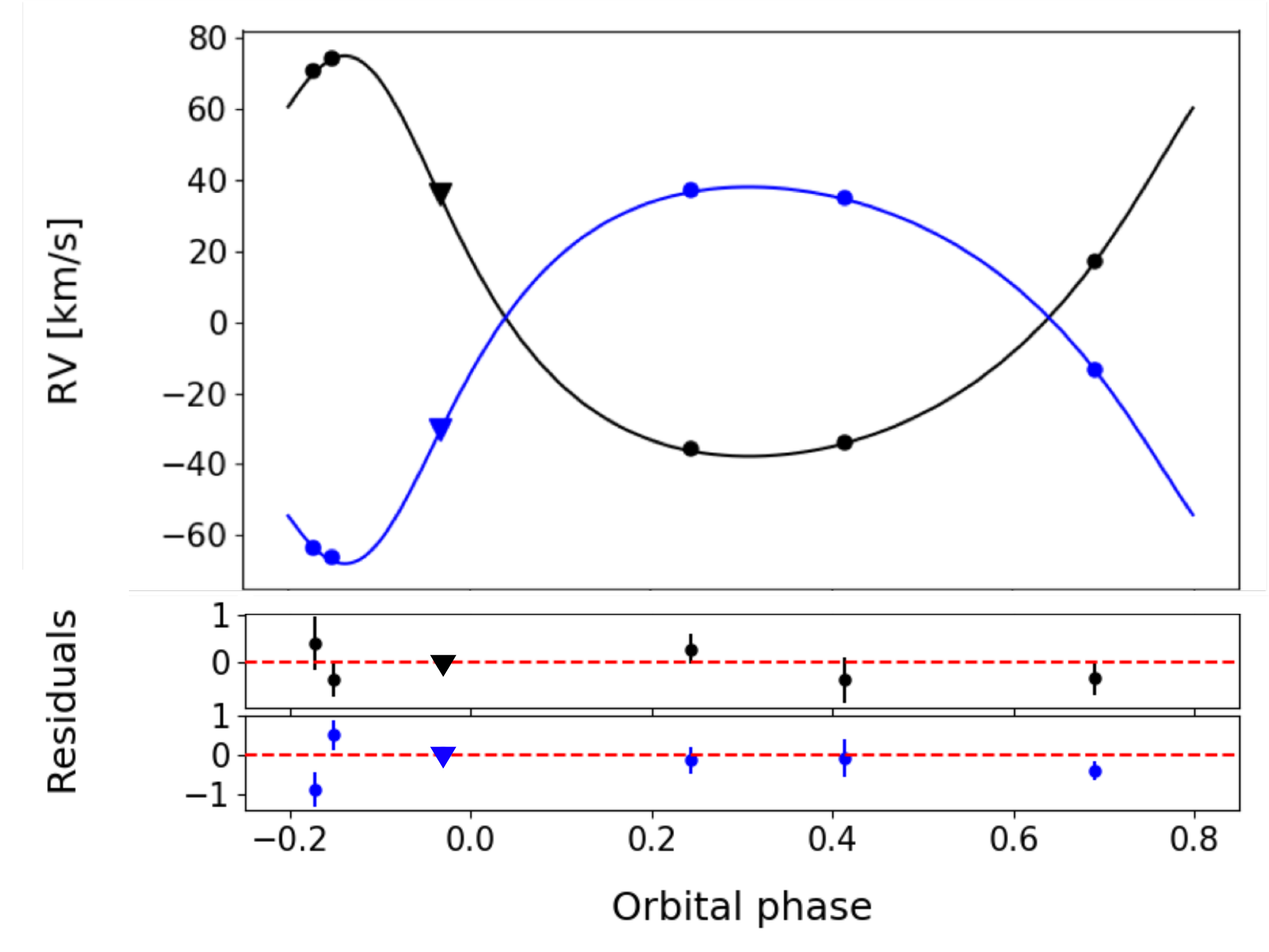}
    \caption{Phase folded radial velocity curve for \NstarA (Black) and \NstarB (Blue), with 5 radial velocity points taken by CORALIE (circles) and a single point obtained with HARPS (triangle). Radial velocities were phase folded on a period of \period\, days. Black and blue lines show the model fit to the radial velocities obtained from global modelling of the system.}
   \label{fig:rv_fit}
\end{figure}

The spectra were cross correlated with an M4 spectral mask and the cross correlation function (CCF) determined in order to derive the radial velocities of both stars. We determined which star each peak belonged to using the well defined ephemeris from the previous photometric observations. We obtained a total of 6 observations spread across a full phase range which proved sufficient to constrain both the orbital eccentricity and masses of the system. The phase folded RV curve fit with our best model is shown in Figure~\ref{fig:rv_fit}. The full radial velocity measurements are given in Table~\ref{tab:RV_summary}.

\begin{table*}
	\centering
	\caption{Radial Velocities for \Nstar A and B.}
	\label{tab:RV_summary}
	\begin{tabular}{cccccc} % six columns, alignment for each
    \hline
BJD$_\mathrm{TDB}$			&	RV		&RV error &	FWHM& 	Contrast& Instrument\\
(-2,450,000)	& (km/s)& (km/s)& (km/s)&(\%) & \\
		\hline
        STAR A& & & & &\\
        \hline
7958.810121	& 36.163  &	0.065&	7.730&	7.1&	HARPS\\
7994.780464	& 17.437  &	0.370&	7.911&  7.4&	CORALIE\\
8670.668149	& -33.780 &	0.490&	6.900&	6.1&    CORALIE\\
8673.822407 & 70.990  &	0.570&	4.600&	2.6&    CORALIE\\
8722.700796 & -35.54  & 0.320&  8.164&  7.0&    CORALIE\\
8742.540179 &74.4587  &   0.350&  8.960&  7.1&    CORALIE \\

		\hline
        STAR B& & & & & \\
        \hline
7958.810121	&-30.007&	0.058&	7.365&	8.6& HARPS\\
7994.780464	&-13.217&	0.250&	7.448&	8.3& CORALIE\\
8670.668149	&-35.000&	0.480&	8.400&	7.1& CORALIE\\
8673.822407	&-63.178&	0.440&	4.700&	3.8& CORALIE\\
8722.700796 &37.210 &   0.330&  8.642&  7.5& CORALIE\\
8742.540179 &-65.785&   0.375&  9.787&  7.6& CORALIE\\
        \hline
 
	\end{tabular}
\end{table*}

\section{Analysis}

\subsection{Spectral Typing}
We obtained near-IR and visible spectra of \Nstar using the X-SHOOTER instrument on ESOs Very Large Telescope (VLT). The observations were taken as part of proposal 0101.C-0181 (Characterising M-dwarf exoplanet hosts and binaries from NGTS, Casewell PI). 

\Nstar was observed on the nights 2018, September 1st and 2nd. Observations were taken using the NIR and VIS arms of X-SHOOTER giving a wavelength coverage of 0.53--2.48~microns. We used an exposure time of 120~s which resulted in a signal to noise ratio of $>$100 for the observations. The spectra were reduced and flux calibrated using the standard ESO X-SHOOTER pipeline \citep{Modigliani2010}.

These spectra were used to determine the spectral type of \Nstar. We first performed the telluric absorption correction on the spectra using ESO's {\sc molecfit} software (\citealt{Smette2015}, \citealt{Kausch2015}). Due to the high signal to noise ratio in our spectrum, we are able to perform these corrections without the need for telluric standard observations. In this case {\sc molecfit} uses a synthetic transmission spectrum determined by a radiative transfer code in order to perform the corrections.

\begin{figure}
	\includegraphics[width=\columnwidth]{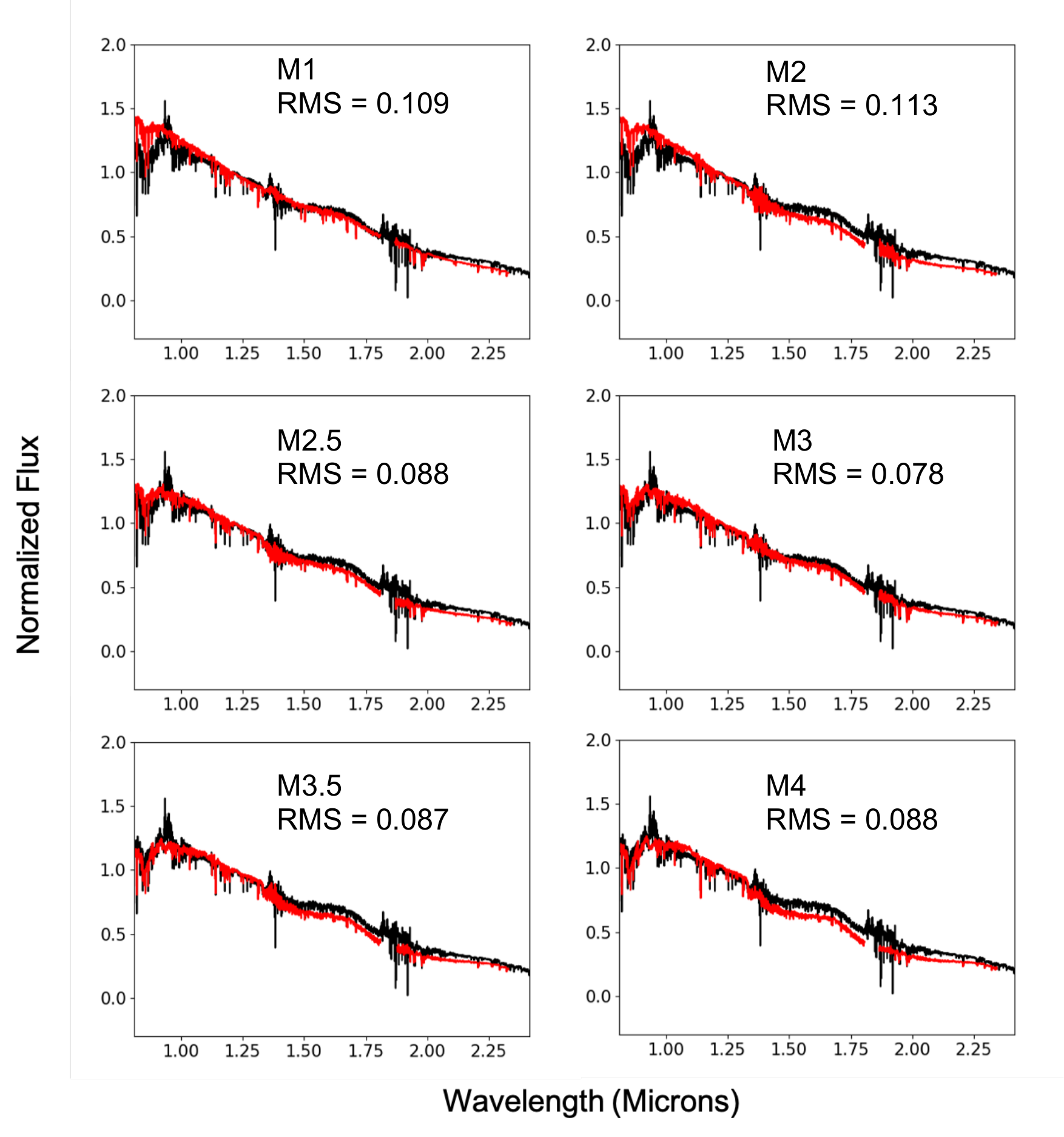}
    \caption{Telluric corrected X-SHOOTER spectrum of \Nstar (black) overlayed with template spectra for a range of M-dwarf spectral types (red). The root-mean-square error for each template is also shown. The best fit to a single spectrum is given by the M3 template.}
   \label{fig:spec_typing}
\end{figure}

The telluric corrected spectra were then normalised at 1.2~$\mu$m  and fit to a variety of M-dwarf spectral templates, minimising the sum of the squares of the difference between the model and real spectra. The best fitting model spectrum was taken as the assumed spectral type of \Nstarns. We used templates of M1 (HD24581), M2 (HD95735), M2.5 (Gl581), M3 (Gl388), M3.5 (Gl273) and M4 (Gl213) as detailed in \citep{Cushing2005} and \citep{Rayner2009}. The best fit was obtained using a template of spectral type M3 (rms 0.078). The fit to each individual template spectrum, and the associated root-mean-square error, is given in Figure \ref{fig:spec_typing}.

\subsection{Global Modeling}\label{sec:global_fitting}
To determine the masses, radii and other parameters for the stars, we performed global fitting of both the photometric data (NGTS, TESS and SAAO) and the RVs (CORALIE, HARPS). This was performed using the eclipsing binary light curve simulation code {\sc ellc} \citep{Maxted2016} in combination with the Markov Chain Monte Carlo (MCMC) sampler {\sc emcee} \citep{Foreman-Mackey2013}. Prior to the global modelling, the raw lightcurves were normalised by their median out of eclipse flux based on the ephemeridies from \textsc{orion}. The NGTS data was binned to 2 minutes to reduce computational time.

We initialized the walkers in a region of parameter space which provided a good initial fit in order to save time. The starting position of each walker was selected from a normal distribution centred on these values. We used the values derived by \textsc{orion} to obtain initial values for both the transit epoch and the orbital period. Initial values for the primary star radius, stellar radius ratio, impact parameter, light ratio and radial velocity components were determined by first running the MCMC for a small number of steps to find values that gave a reasonable initial fit.

Limb darkening parameters were obtained using the \textsc{ldtk} software \citep{Parvianen2015}. A quadratic limb darkening law was used with stellar properties, e.g., \teff\, and \logg\, estimated based on the previously determined spectral type. Limb darkening coefficients and uncertainties were calculated directly with \textsc{ldtk}, for each photometric filter used, and placed as priors for the fitting process.

 As we were unsure of the level of eccentricity of the system, although we expect it to be reasonably high given the position of the secondary eclipse, we initially set it to zero, and allowed the sampler to determine the level of eccentricity. We also incorporated a radial velocity jitter term in our modelling to account for stellar noise, as well as normalisation scaling parameters for each of the three lightcurves.

We used 200 walkers and 45000 steps to model the lightcurve using the \textsc{emcee} sampler. Each walker used initial parameters that were randomized in a Gaussian ball around the values previously determined to give a good initial fit. We note that alterations to our initial values did not preclude the ability of the model to obtain a good fit, but did increase the burn in time required to do so. 10000 steps were discarded as burn-in and not used when analysing the results of the modelling.
 
\begin{figure*}
   \centering
  \includegraphics[width=\textwidth]{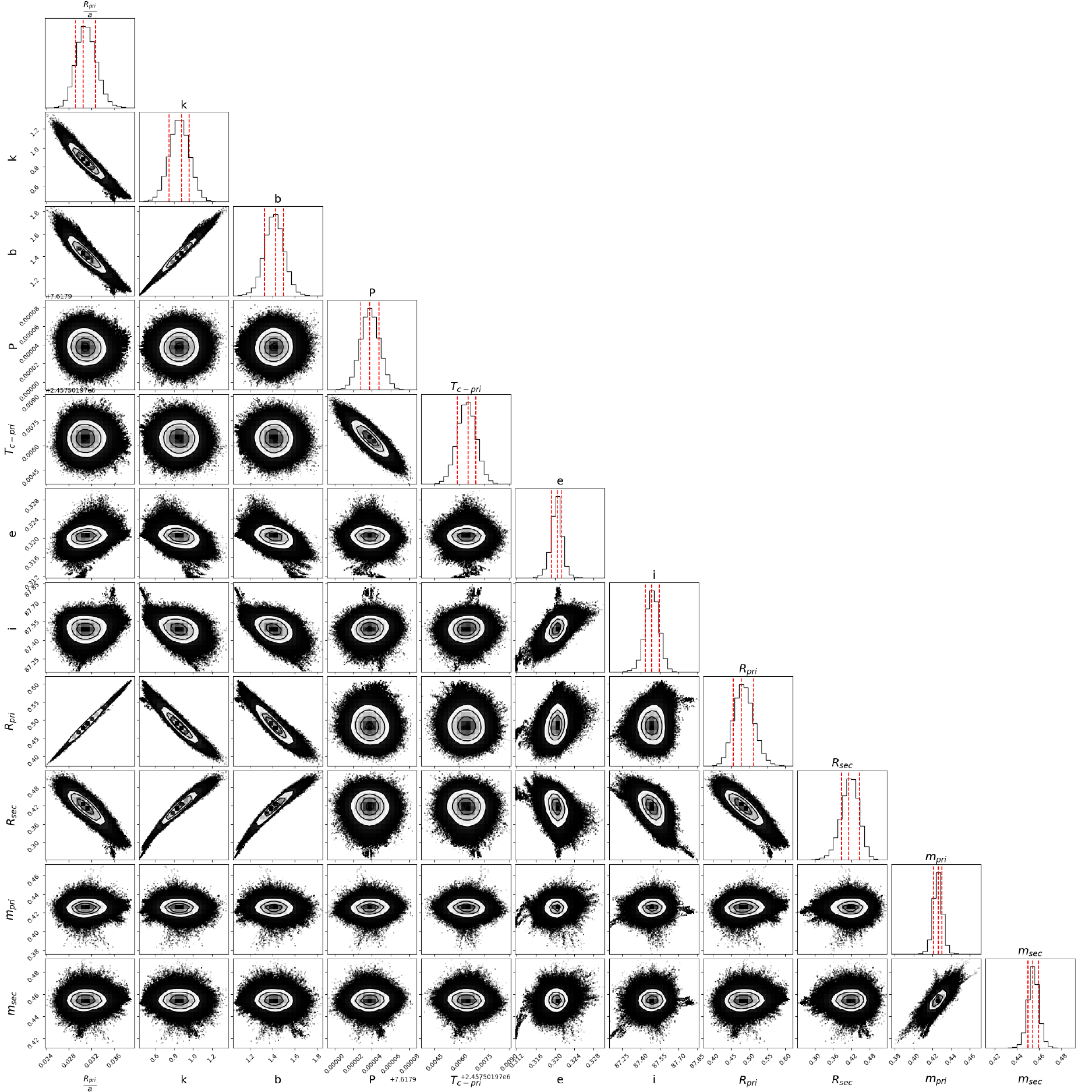}
   \caption{Corner plot showing the distribution of some key orbital parameters obtained from MCMC modelling of the system. Descriptions of each parameter are given in Table \ref{tab:fit_params}}
   \label{fig:corner}
\end{figure*}

 The parameters which obtain the best model fit to the overall data set are given in Table \ref{tab:fit_params}, and a corner plot showing the convergence of some key parameters is shown in Figure \ref{fig:corner}. Our modelling gives best fit masses of $M_{\rm A}$=\NstarMassA$M_{\odot}$, $M_{\rm B}$=\NstarMassB $M_{\odot}$ and radii of $R_{\rm A}$=\NstarRadA $R_{\odot}$, $R_{\rm B}$=\NstarRadB $R_{\odot}$\, suggesting a binary consisting of two roughly equal sized early-mid M-dwarf stars. The stars are in a very grazing eclipse, with an inclination of \incl \textdegree ~ and impact parameter of \impactparam. This leads to a sizeable uncertainty in the radii of the stars. Our global modelling also confirms the eccentricity of the system as suggested by both the photometry and radial velocity measurements, with a value of \ecc. This leads to a pericentre distance of \pericentre\ $R_{\odot}$ for the orbit.

\begin{table*}
\renewcommand{\arraystretch}{1.35}
	\centering
	\caption{Fitted global modelling parameters and subsequent derived parameters for the \Nstar system. The modal values of the posterior distributions were adopted as the most probable parameters, with the 68.3 percent (1$\sigma$) highest probability density interval as the error estimates.}
	\label{tab:fit_params}
	\begin{tabular}{ccccc} % six columns, alignment for each
    \hline
Quantity & Description  & unit & Value & Error\\
\hline
Fitted parameters \\ 
{$\frac{R_{\rm pri}}{a}$} & radius ratio of primary to semi-major axis & none & $0.02952$ & $^{+0.00241}_{-0.00159}$\\

$k$ & radius ratio of stars, $R_{\rm sec}/R_{\rm pri}$ & none & $0.89166$ & $^{+0.10813}_{-0.14293}$\\

$b$ & impact parameter, $\cos{(i)}/R_{\rm pri}$ & none & $1.43083$ & $^{+0.08823}_{-0.11604}$\\

$\sqrt{e} \cos \omega$ & orbital eccentricity and argument of periastron term & none & $0.54630$ & $^{+0.00117}_{-0.00141}$\\

$\sqrt{e} \sin \omega$ & orbital eccentricity and argument of periastron term & none & $0.14821$ & $^{+0.00926}_{-0.00819}$ \\

$P$ & orbital period & days & $7.61793$ & $^{+0.00000516}_{-0.00000544}$ \\

$T_{c}$ & epoch of primary eclipse centre & BJD & $2457501.97719$ & $^{+0.00040}_{-0.00039}$ \\

{$\sigma_{\rm NGTS}$} & systematic error in NGTS light curve & norm. flux & $0.00638$ & $^{+0.0000638}_{-0.0000627}$\\

{$\sigma_{z}$} & systematic error in z' light curve & norm. flux & $0.000232$ & $^{+0.000260}_{-0.000163}$ \\

{$\sigma_{\rm TESS}$} & systematic error in TESS light curve & norm. flux & $0.00283$ & $^{+0.000120}_{-0.000121}$ \\

{$\beta_{\rm NGTS}$} & normalised flux scale factor in NGTS data & none & $1.00019$ & $^{+0.0000641}_{-0.0000655}$ \\

{$\beta_{z}$} & normalised flux scale factor in z' data & none & $0.99913$ & $^{+0.00030}_{-0.00030}$  \\

{$\beta_{\rm TESS}$} & normalised flux scale factor in TESS data & none & $0.99996$ & $^{+0.0000598}_{-0.0000635}$  \\

$u_{\rm{NGTS-pri}}$ & linear LDC of primary in NGTS band & none & $0.48571$ & $^{+0.03281}_{-0.03237}$ \\

$u'_{\rm{NGTS-pri}}$ & quadratic LDC of primary in NGTS band & none & $0.30374$ & $^{+0.05351}_{-0.05403}$ \\

$u_{\rm{NGTS-sec}}$ & linear LDC of secondary in NGTS band & none & $0.51173$ & $^{+0.03312}_{-0.03437}$ \\

$u'_{\rm{NGTS-sec}}$ & quadratic LDC of secondary in NGTS band & none & $0.36833$ & $^{+0.05681}_{-0.06064}$  \\

$u_{\rm{z-pri}}$ & linear LDC of primary in z' band & none & $0.34917$ & $^{+0.03633}_{-0.03647}$ \\

$u'_{\rm{z-pri}}$ & quadratic LDC of primary in z' band & none & $0.31609$ & $^{+0.06549}_{-0.06555}$ \\

$u_{\rm{z-sec}}$ & linear LDC of secondary in z' band & none & $0.35603$ & $^{+0.03679}_{-0.03693}$ \\

$u'_{\rm{z-sec}}$ & quadratic LDC of secondary in z' band & none & $0.33512$ & $^{+0.06636}_{-0.06834}$ \\

$u_{\rm{TESS-pri}}$ & linear LDC of primary in TESS band & none & $0.46757$ & $^{+0.03252}_{-0.03243}$ \\

$u'_{\rm{TESS-pri}}$ & quadratic LDC of primary in TESS band & none & $0.28081$ & $^{+0.05562}_{-0.05439}$ \\

$u_{\rm{TESS-sec}}$ & linear LDC of secondary in TESS band & none & $0.45157$ & $^{+0.03278}_{-0.03332}$ \\

$u'_{\rm{TESS-sec}}$ & quadratic LDC of secondary in TESS band & none & $0.22835$ & $^{+0.05651}_{-0.05540}$ \\

$J_{\rm NGTS}$ & light ratio in NGTS band & none & $0.85668$ & $^{+0.23697}_{-0.29601}$  \\

$J_{z}$ & light ratio in z' band & none & $0.68200$ & $^{+0.20077}_{-0.24757}$  \\

$J_{\rm TESS}$ & light ratio in TESS band & none & $0.77850$ & $^{+0.20418}_{-0.24760}$  \\

{$K_{\rm pri}$} & radial velocity semi-amplitude of primary & km/s & $56.48061$ & $^{+0.30451}_{-0.28743}$  \\

{$K_{\rm sec}$} & radial velocity semi-amplitude of secondary & km/s & $52.88555$ & $^{+0.30385}_{-0.27938}$  \\

{$\Gamma_{\rm pri}$} & systemic velocity measured from primary & km/s & $1.84146$ & $^{+0.22779}_{-0.24778}$ \\

{$\Gamma_{\rm sec}$} & systemic velocity measured from secondary & km/s & $1.94810$ & $^{+0.21990}_{-0.22469}$ \\

{$\sigma_{\rm RV}$} & jitter in RV data & km/s & $0.29331$ & $^{+0.17764}_{-0.32047}$  \\

%\cmidrule(lr){1-5}

Derived parameters \\
%\cmidrule(lr){1-5}

{$R_{\rm pri}$} & radius of primary & $R_\odot$ & $0.46111$ & $^{+0.03795}_{-0.02492}$ \\

{$R_{\rm sec}$} & radius of secondary & $R_\odot$ & $0.41060$ & $^{+0.02670}_{-0.03880}$ \\

$m_{\rm pri}$ & mass of primary & $M_\odot$ & $0.42559$ & $^{+0.00558}_{-0.00491}$  \\

$m_{\rm sec}$ & mass of secondary & $M_\odot$ & $0.45451$ & $^{+0.00583}_{-0.00517}$ \\

{$a$} & semi-major axis of system & $R_\odot$ & $15.61805$ & $^{+0.06457}_{-0.05644}$ \\

$i$ & orbital inclination & $deg$ & $87.58709$ & $^{+0.05265}_{-0.04650}$ \\

$e$ & orbital eccentricity & none & $0.32034$ & $^{+0.00120}_{-0.00117}$ \\

$\log\,g_{\rm pri}$ & primary surface gravity & $cm s^{-2}$ & $4.73846$ & $^{+0.04797}_{-0.0.06808}$\\

$\log\,g_{\rm sec}$ & secondary surface gravity & $cm s^{-2}$ & $4.86771$ & $^{+0.08721}_{-0.05487}$ \\

$T_{\rm 14-pri}$ & primary eclipse duration & hours & $2.28084$ & $^{+0.02831}_{-0.04195}$ \\

$T_{\rm 14-sec}$ & secondary eclipse duration & hours &   $2.30355$ & $^{+0.04217}_{-0.02987}$ \\

	\hline
    \end{tabular}
\end{table*}

\subsection{Stellar Rotation}\label{sec:rot}

\begin{figure}
   \centering
	\includegraphics[width=\columnwidth]{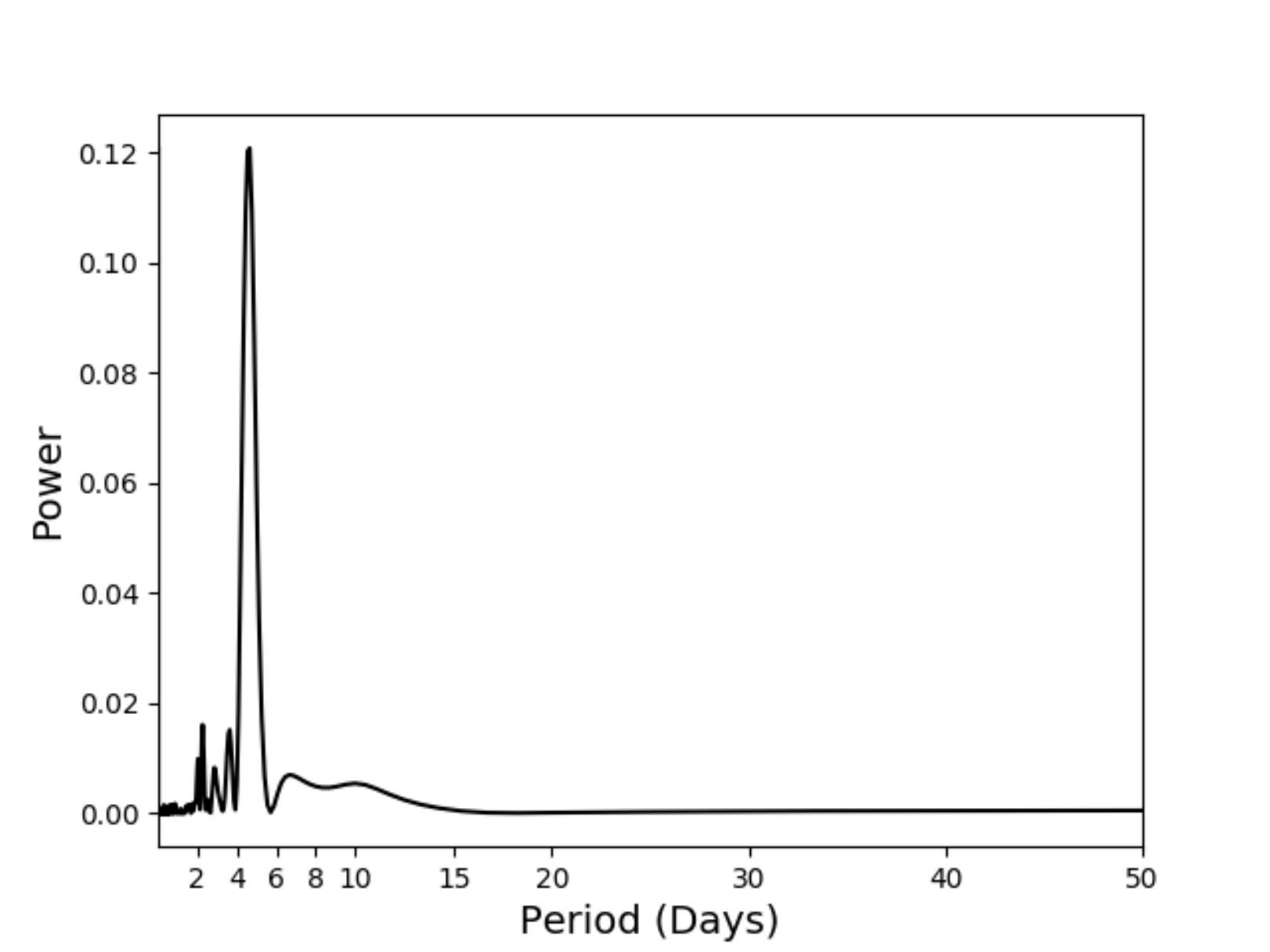}
    \caption{Lomb-Scargle periodogram for the TESS lightcurve after the removal of primary and secondary eclipse events. Note the strong peak at 4.647 days as a possible signature of stellar rotation.}
   \label{fig:periodogram}
\end{figure}

\begin{figure}
   \centering
	\includegraphics[width=\columnwidth]{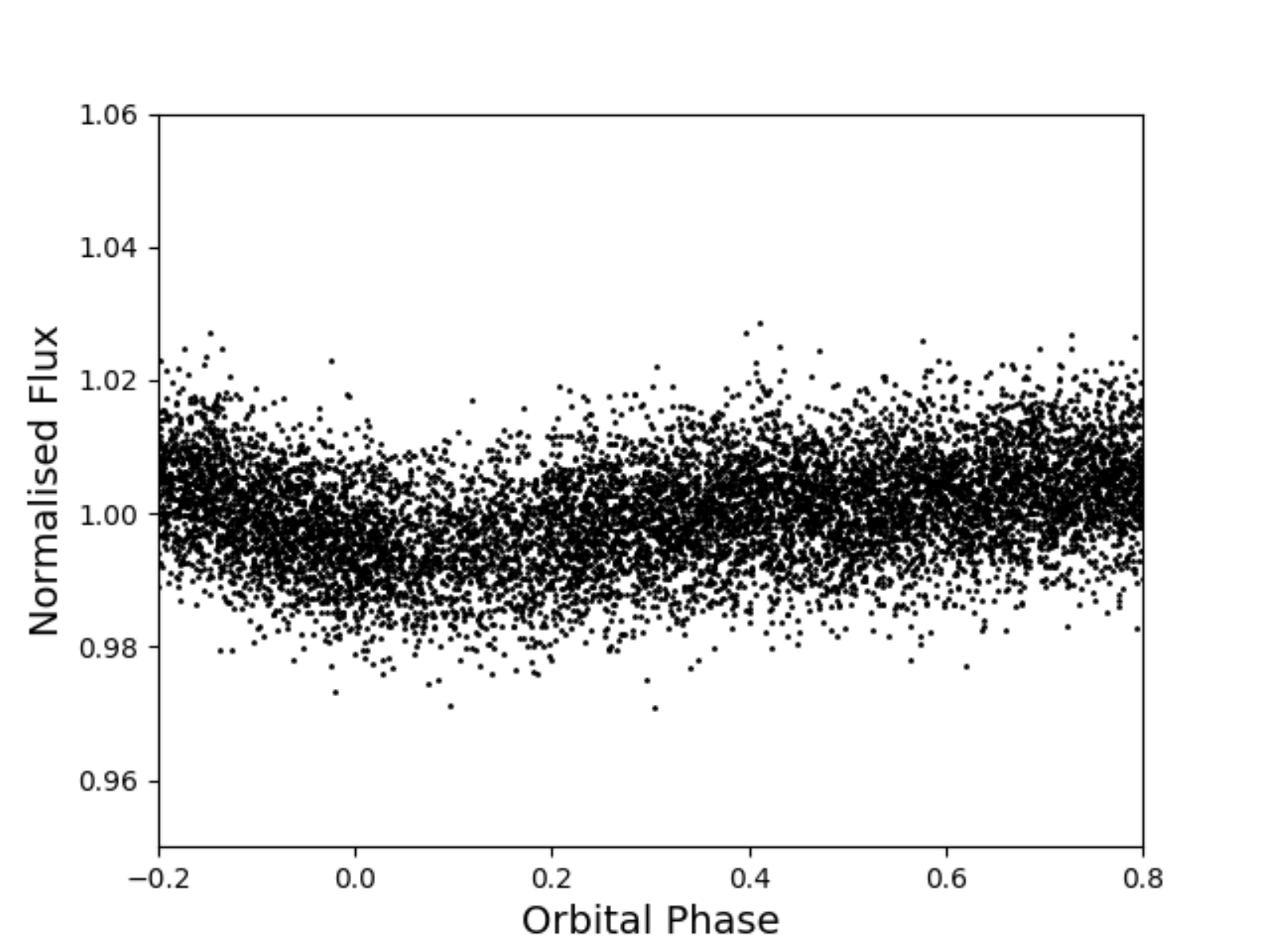}
    \caption{TESS Lightcurve phase folded on a period of 4.647 days. Clear Modulation can be seen with an amplitude of around 1\%,  likely a signature of the rotation period of one (or both) of the stars}
   \label{fig:rotation_lcs}
\end{figure}

In deriving the system parameters we also attempted to determine rotation periods for the stars. From inspection of the TESS lightcurve, we see a clear sinusoidal like variation which could be related to the rotation period of at least one of the stars. In order to fully characterise this, we ran a standard Lomb-Scargle period search to identify any periodic varitations in the photometry. Prior to performing this we removed the system's primary and secondary eclipses from the photometry, to prevent periods based on these events from being detected.

We performed the Lomb-Scargle search on both the NGTS and TESS data. The periodogram for the TESS lightcurve is shown in Figure \ref{fig:periodogram}. A clear peak can be seen with an associated period of 4.647 days. The NGTS lightcurve also shows similar periodic variability, with an identified period of 4.595 days. The identification of similar periodic variability in two independent sets of data is a strong indicator of validity.

The TESS lightcurve, folded on this suspected rotation period, is shown in Figure \ref{fig:rotation_lcs}. A clear sinusoidal variation at  $\sim 1\%$ level can be seen, which is likely associated with the rotation period of at least one of the stars. The NGTS lightcurve also shows some sinusoidal modulation, however this is less noticeable than in the TESS data, due to the increased scatter of the data.

 \subsection{SED Fitting}
In order to determine the effective temperatures of both stars we fitted their Spectral Energy Distributions (SEDs), following a method similar to \citet{Gillen2017}. We fit both stars simultaneously using the PHOENIX v2 models \citep[][]{Husser13}. The PHOENIX models were initially convolved with our catalogue photometry to generate a grid of fluxes in \teff\ -- \logg\ space. This grid was then used to model the combined photometry of both stars. We fit for \teff, \logg\ and the radius $R$ of each component, the distance $D$, along with an error inflation term, $\sigma$, which is used to account for underestimated catalogue uncertainties. To fully explore the posterior parameter space we performed our fitting using a MCMC process using \textsc{emcee} \citep[][]{Foreman-Mackey2013}. We used 200 walkers for 50,000 steps and used the final 10,000 to sample the posterior distribution. During fitting we applied priors on the system distance and the radius of each component. We used a Gaussian prior for the system distance, using the \Gaia\ DR2 measured value from \citet{BailerJones18}. For the radius of each component we used the measured values and uncertainties from our global fitting in \S\ref{sec:global_fitting}, using two half-Gaussians for the prior distribution.

\begin{figure}
   \centering
	\includegraphics[width=\columnwidth]{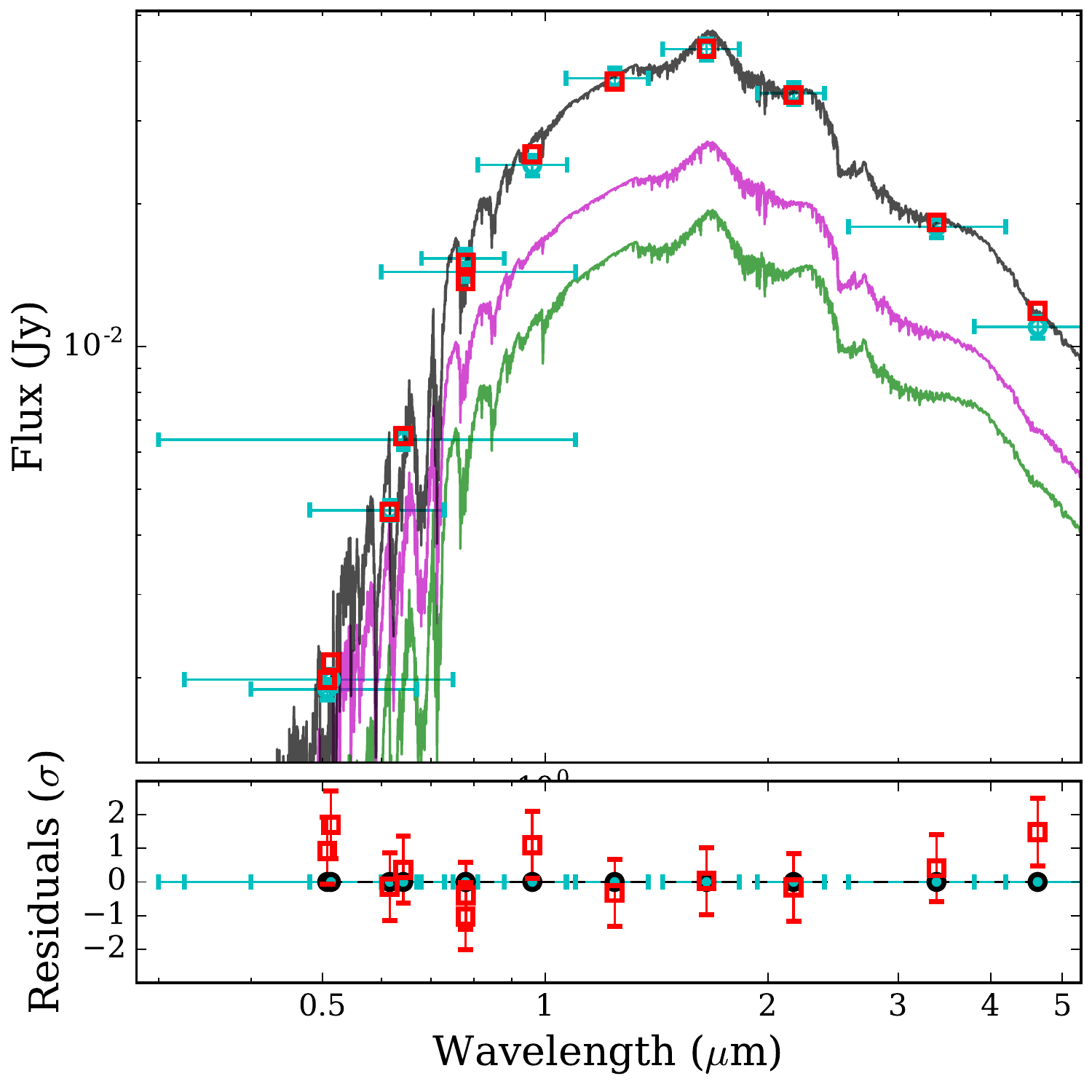}
    \caption{Upper panel -- spectral energy distribution of \Nstar\, fit with a combination of PHOENIX~v2 model spectra (black line). The magenta and green lines show the individual model spectra for each star, the blue points indicate the SED of the combined system. Lower panel -- fit residuals for the combined spectrum.}
   \label{fig:SED}
\end{figure}

The fit to the system SED is shown in Figure~\ref{fig:SED}. From this we derived effective temperatures of \teff=\TeffA$K$ and \teff=\TeffB$K$ for the primary and secondary stars respectively. This is in agreement with the temperature expected for two M3 Dwarfs, and is consistent with the mass and radius derived from the global modelling of the system.

\section{Discussion}

\subsection{Mass Radius relation}
\begin{figure}
   \centering
	\includegraphics[width=\columnwidth]{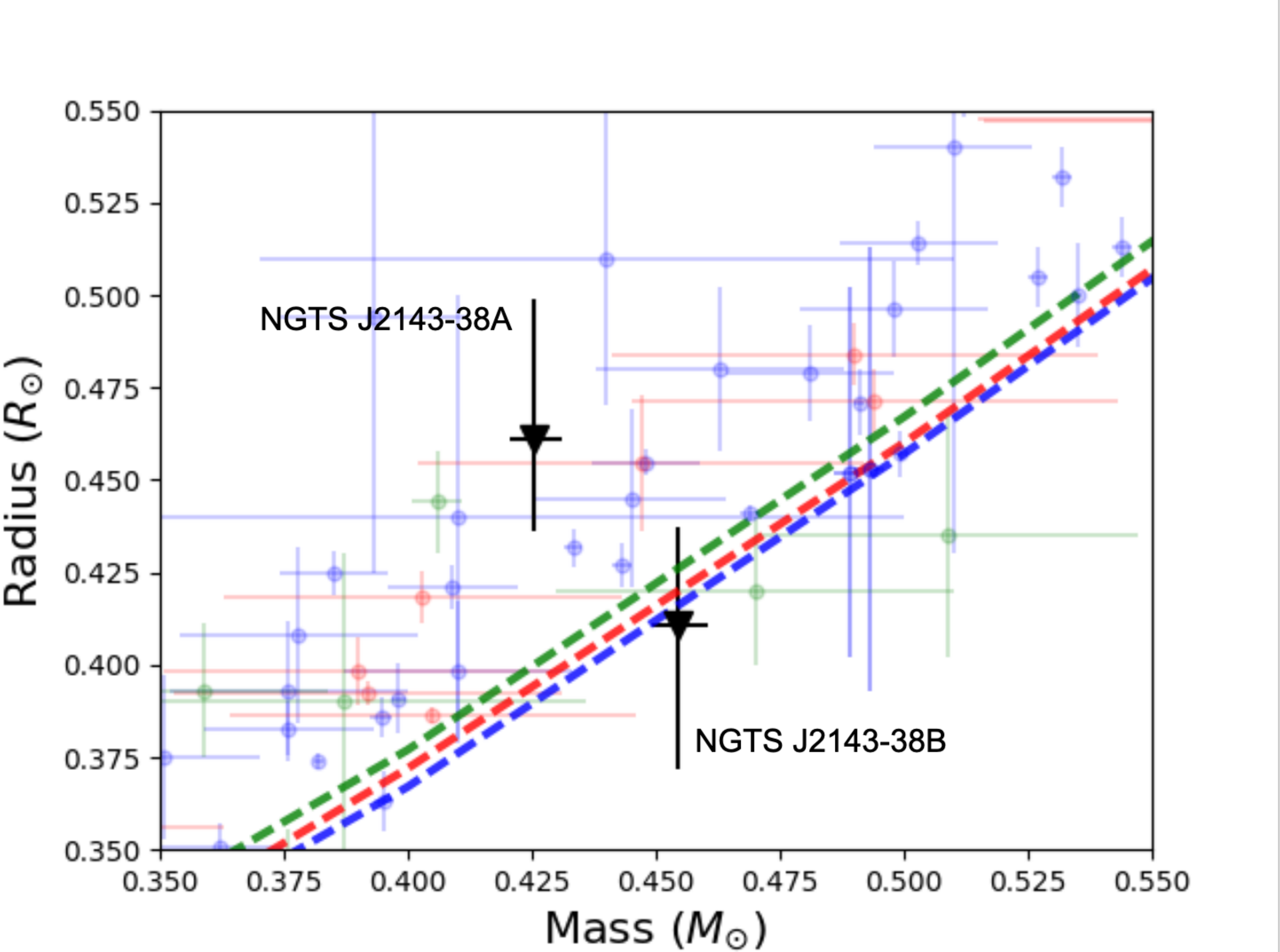}
    \caption{Comparison between the two components of \Nstar and the model stellar isochrones from \citealt{Baraffe2015}. A 0.5-Gyr isochrone is plotted in blue,  a 1-Gyr isochrone is plotted in red, and a 5-Gyr isochrone is plotted in green. \NstarA and \NstarB are indicated by black triangles. Similar M-dwarfs from \citealt{Parsons2018} are also plotted. Blue points are double lined M-dwarf binaries, red points are single stars and green points are M-dwarf secondaries in binaries with other spectral types.}
   \label{fig:massradiuszoom}
\end{figure}

%\begin{figure}
 %  \centering
%%   \caption{Mass Radius relation for well characterised M-dwarfs from \citealt{Parsons2018}. \NstarA and \NstarB are indicated by black triangles. The theoretical track from \citealt{Baraffe2015} is shown as the dashed blue line. Blue points are double lined M-dwarf binaries, red points are single stars and green points are M-dwarf secondaries in binaries with other spectral types.}
 % \label{fig:massradius}
%\end{figure}

Since \Nstar\, is in a grazing eclipse, there is a greater uncertainty in the masses and radii of the stars than there would be if the eclipses were to occur across the full face of each star. Nonetheless, as the stars lie in a parameter space with a relatively high number of measurements, we are able to compare our results to a large number of known systems, as well as to theoretical model predictions, to see where our stars fall amongst the already known significant scatter of the M-dwarf mass--radius relation.

Figure \ref{fig:massradiuszoom} shows the mass and radius of the two stars plotted against model stellar isochrones from \cite{Baraffe2015}. \NstarB lies within 1 sigma of model stellar isochrones, in agreement with model predictions for mass and radius. \NstarA however, appears to be inflated relative to its mass, differing from the model prediction for radius by almost 20\%. 

%\lr{How many sigma is Star A away? Given upper uncertainties, it looks like it could be within the top 5 most outlying stars?}

When these stars are compared with the wider population of M-dwarfs, e.g., Figure~\ref{fig:massradiuszoom}, we see that they lie within the expected scatter of M-dwarf uncertainty relative to theoretical relationships. However, the inflation of \NstarA is still clearly visible, indicating that it is among the most inflated M-dwarfs known. From the known M-dwarfs shown in \ref{fig:massradiuszoom}, it is apparent that the majority of outliers are also early--mid type stars in M-M binary systems, with a configuration similar to \Nstarns. As one of the most inflated systems, \Nstar\, represents an excellent opportunity for investigating the underlying inflation mechanisms  responsible for the oversizing observed in some M-dwarfs, and  currently missing from physical models.

\cite{Boyajian2012} show that stars in eclipsing binaries have larger radii than single stars of the same temperature. This may be an explanation for the inflation seen in \NstarA. It is perhaps slightly unusual however that only one of the stars shows this inflation, with \NstarB showing reasonable agreement with model predictions. However, \cite{Kraus2017} report the discovery of an M-dwarf binary in which the primary star matches theoretical predictions whereas the secondary star is oversized relative to predictions. This is the opposite scenario to our system, in which the primary is inflated and the secondary is consistent with model predictions, which further highlights the complicated inconsistencies between models and predictions.

We note that while convention dictates that the primary star is defined to be that which gives the deepest eclipse when occulted by its companion (as in \citealt{Gillen2017}), this does not necessarily imply that the primary star is more massive than the secondary. Indeed, for \Nstar\, we find the secondary star to be slightly larger in mass than the primary.

\subsection{Mass Effective Temperature Relationship}

\begin{figure}
   \centering
	\includegraphics[width=\columnwidth]{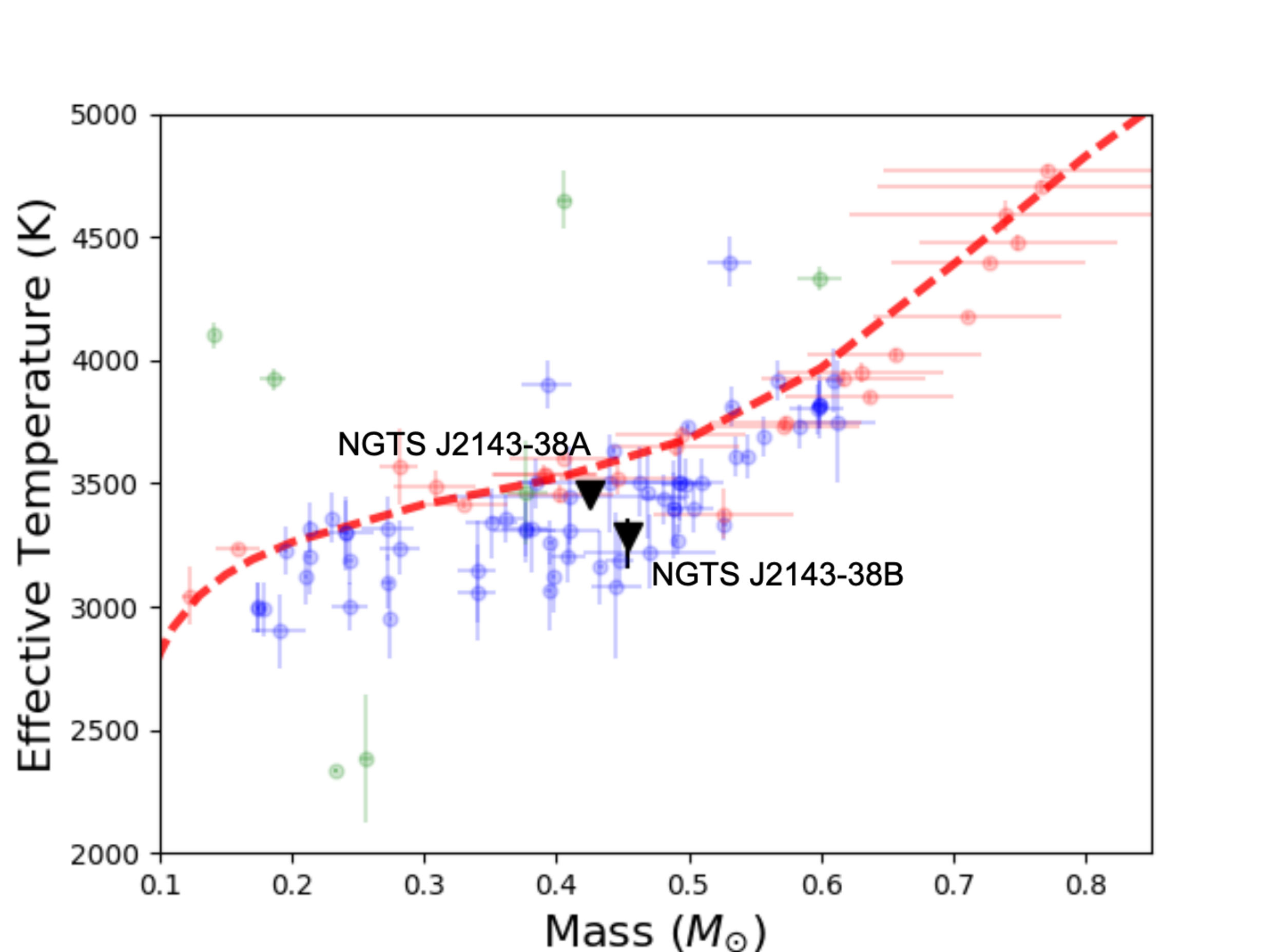} 
    \caption{Mass temperature relation for the sample of M-Dwarfs given in \citet{Parsons2018}. The components of \Nstar\, are plotted in black, and a 5~Gyr isochrone is plotted in red for comparison. Blue points are double lined M-Dwarf Binaries, red points are single stars and green points are M-Dwarf secondaries in binaries with other spectral types.}
   \label{fig:masstemp}
\end{figure}

We compare the effective temperatures for both stars derived in \S\ref{sec:global_fitting} to the known temperatures of the sample of M-Dwarfs listed in \cite{Parsons2018}. This is shown in Figure~\ref{fig:masstemp}, along with theoretical tracks \citep{Baraffe2015} for stars of this mass range.

Like similar stars in this mass regime (see e.g., \citealt{Lopez2007}) both of our stars are cooler than model predictions by   $\approx100K$. This is consistent with the trend seen in partially convective stars, where luminosity (and hence temperature) is not strongly controlled by the conditions in their outer layers. Rather, the temperatures for these stars are affected by the conditions in the core. As seen in Figure~\ref{fig:masstemp}, lower mass (and hence fully convective) stars tend to agree better with model predictions, due to the fact that their luminosities are strongly determined by the conditions in the outer layers of the star.

\subsection{Eccentricity}

\begin{figure}
   \centering
	\includegraphics[width=\columnwidth]{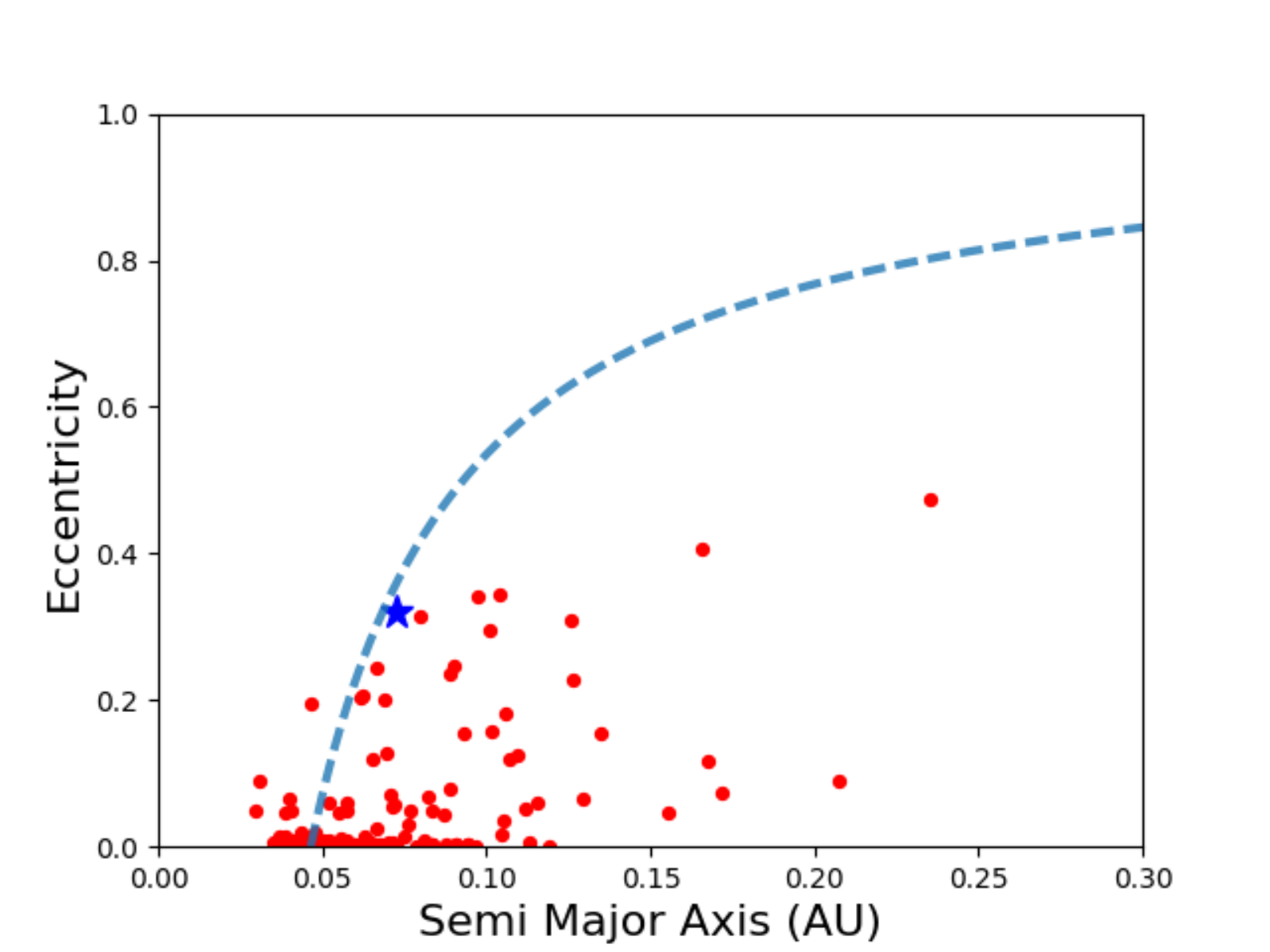}
    \caption{Plot illustrating the relationship between the semi major axis of the binary and the eccentricity of the binary orbit for a sample of low mass eclipsing binaries described in \citealt{Triaud2017}. \Nstar is indicated in blue. A line of constant pericentre distance (10 $R_{\odot}$) is shown in blue for comparison.}
   \label{fig:ecc_binaries}
\end{figure}

The global modelling of the system gives a value for the eccentricity of \ecc. This is unusual for M-dwarf binary systems which are often found in tight, almost perfectly circular orbits. Most double-lined eclipsing M-dwarf binaries with similar periods (such as \citealt{Kraus2017, Orosz2012}) show no such eccentricity in their orbits. LPSM J1112+7626 \citep{Irwin2011} is an M-dwarf binary system with an eccentricity of 0.24, however this system has a much longer orbital period (41.032 days) and as such a large level of eccentricity is less unusual.

There are some examples of eclipsing binaries with similar orbital periods that do exhibit some eccentricity. \cite{Triaud2017} obtained spectroscopic orbits of 100 low mass eclipsing binaries with a range of orbital periods. Figure \ref{fig:ecc_binaries} shows the eccentricity of these binary systems as a function of semi-major axis. Whilst there are clearly examples of eccentricity in binary orbits with periods on the order of days, there are no examples of shorter orbital period binaries with greater eccentricity than \Nstar. It is also important to note, that the objects in \cite{Triaud2017} are F, G and K type stars with M-dwarf secondary companions, and therefore do differ in nature to double M-dwarf binaries like \Nstar. Of the double-lined M-Dwarf binaries listed by \citet{Parsons2018}, none have a greater eccentricity than \Nstar, regardless of the period or semi major axis of the systems.

\subsubsection{Tidal Circularisation}

A possible explanation for this eccentricity may be that the system is rather young ($<<$1 Gyr). Whilst tidal evolution of the binary may act to reduce the eccentricity of the orbit, there is no reason to expect the system to have began with little, or zero, eccentricity. If the system is sufficiently young that the eccentricity is yet to be reduced by these effects then this may explain the presence of such a large eccentricity relative to the period of the orbit.

If this system is not young, then it is possible that tidal effects which would normally dampen the eccentricity of the system are not strong enough to do so in this scenario. We have already shown (in Figure \ref{fig:ecc_binaries}) that relative to its semi-major axis, \Nstar\ has an extremely eccentric orbit. This makes the system a potentially useful test case for investigating the strength of tidal circularisation effects in binary star systems.

The circularisation timescale of a binary system can be defined as follows \citep{Hurley2002}.

\begin{equation}
\centering
    \frac{1}{\tau_{circ}}=\frac{21}{2}\frac{k}{T}q_2 \left( 1+q_2 \right) \left(\frac{R_{pri}}{a} \right)^8 \; .
\end{equation}{}

\noindent Here $q_2$ is the mass ratio $M_2$/$M_1$, $k$ and $T$ are constants which depend on the structure of the stars and the timescale on which significant changes can happen to the orbit. For convenience, we can express this in terms of the stellar tidal quality factor, $Q_{*}$, such that the circularisation timescale is given by:

\begin{equation}
\centering
    \tau_{circ}=\frac{2}{21}Q_{*}\frac{1}{q_2(1+q_2)}\left(\frac{a}{R_{pri}}\right)^8  \; .
\end{equation}{}

\noindent The stellar tidal quality factor, $Q_{*}$ dictates the strength, or efficiency of tidal forces in a given system. This factor has been calculated for a range of systems, including solar system bodies such as Jupiter, however the theoretical processes that determine this quantity are not well understood, and consequently values for $Q_{*}$ vary greatly. 

Adopting a $Q_{*}$ value of $10^6$ as has been assumed for some eclipsing binary and transiting exoplanet systems, gives a circularisation timescale of 1.09~Gyr. As \Nstar\, is in a highly eccentric orbit, then it must clearly be younger than this, or its orbit would have had adequate time to circularise. However previous studies cited by \cite{Penev2012} suggest that $Q_{*}$ could range anywhere between $10^5$ and $10^9$, and thus it remains extremely difficult to constrain the age of the system in this way. Thus, it is clear that either the system is significantly younger than 1.09 Gyr or alternatively $Q_{*}$ $>>$ $10^6$. If the age of the system were able to be determined then tighter constrains on a value of $Q_{*}$ could be placed.

The age of this system is difficult to determine accurately. Stellar rotation rates from which the age can be inferred, are difficult to determine for binary stars. Stellar activity can also be a useful indicator of age, as stars tend to be more active (and thus exhibit more flares) earlier in their evolution. We searched the light curve of \Nstarns, which contained 139 Nights of data spanning $\approx$ 7~months, and identified only two clear flare events in this time interval. This is likely just an indicator of the inherent activity of M-Dwarf stars rather than a strong indicator of age. We also note that we examined the spectra of the system for Lithium absorption, a signature of young stars, and find no evidence of this. Additionally the star does not appear to be part of a known moving group, from which its age could then be determined.  

\cite{Zahn1989} calculated a theoretical cutoff period, $P_c$ below which any binary system should have a circular orbit on the main sequence. For two stars with masses of 0.5$M_{\odot}$ with an initial eccentricity of 0.3, they give a cutoff period of 7.28 days. The components of \Nstar are slightly less massive than this, and have a longer period, indicating that we may not expect this system to have been tidally circularised. Their determination that shorter period, more eccentric systems should not exist is supported by the fact that our system is of a longer period. They also determine that most tidal circularisation occurs before the star joins the main sequence, and thus if our system is eccentric due to its age, then it must be very young.

\subsubsection{Tidal Synchronisation}
Additionally to investigating the tidal circularisation of the system, insight can be gained by considering the tidal synchronisation of the system. As a binary star system evolves, the spin of the stars tends to be synchronised with the orbit due to tidal effects. The timescale on which this occurs is much shorter than the timescale required to circularise the orbit \citep{Hurley2002}, and thus investigating this effect could provide some insight into the age and evolution of the system providing we are able to determine a rotation rate for either of the stars.

In \S\ref{sec:global_fitting} we showed that there were periodic signals of 4.595 days and 4.647 days from the NGTS and TESS data respectively, which are likely related to the rotation period of at least one of the stars. Both of these periods, if they indeed correspond to the rotation of the stars, are significantly different from the orbital period of the binary. This implies that the system is not yet tidally spin-orbit synchronised. This would further the possibility that the system may be in the early stages of tidal evolution and thus the high level of eccentricity detected may not be entirely unexpected.

\subsubsection{Possible Tertiary Companion}

\cite{Anderson2017} show that eccentricity in binary systems with periods on the order of days can be induced by the presence of a hidden tertiary companion to the binary. Assuming the inclination between the orbits of the inner binary and the outer companion is sufficiently high, it is possible for the eccentricity to undergo periodic excursions to large values. These are known as Lidov-Kozai (LK) cycles (\citealt{Lidov1962,Kozai1962}).

It is believed that binaries with orbital periods shorter than around 7 days are not primordial, and that their current orbital configuration has evolved from a wider configuration perhaps due to these LK cycles (\cite{Mazeh1979}, \cite{Eggleton2001}, \cite{Fabracky2007}, \cite{Naoz2014}). It is indeed possible that this same interaction could induce the eccentricity that is seen in \Nstar. \cite{Tokovinin2006} observed a wide sample of spectroscopic binaries to search for higher order multiplicity. They found the presence of a hierarchical triple to be strongly dependent on the period of the binary orbit, with this being around 50 per cent for a 7 day period of the inner binary, as in our scenario. Therefore it is a reasonable possibility that a tertiary component may be present in this system.

It is important to note however, that the presence of a tertiary star does not guarantee that the binary eccentricity will be excited by these effects. It is strongly dependent on other factors, such as the orbital separation of the inner binary and the inclination between the orbits of the inner binary system and that of the tertiary star. Thus we cannot conclude that \Nstar has been raised to its eccentric orbit by these effects. We also note that there are no nearby Gaia DR2 sources in the vicinity of this object. The closest Gaia source is around 20" away, and is clearly not associated with this system based on its parallax and proper motion. From this we can infer that any tertiary star is either too faint for Gaia, or is too blended with  \Nstar to be resolved.

\section{Conclusions}
We report the discovery of an M-dwarf binary system \Nstar\
using photometry from the Next Generation Transit Survey. Follow-up observations allowed us to determine the masses and radii of the two stars, indicating a pair of roughly equal sized M3 stars.  Global modelling confirmed the eccentric nature of the system, and showed it to be the most eccentric M-Dwarf binary known, and one of the most eccentric binaries of any type relative to its semi-major axis. Additional analysis of the system, including an accurate determination of the age, could provide interesting insights into the effects of tidal circularisation in binary star systems. 

\section*{Acknowledgements}
Based on data collected under the NGTS project at the ESO La Silla Paranal Observatory. The NGTS facility is operated by the consortium institutes with support from the UK Science and Technology Facilities Council (STFC) project ST/M001962/1. This paper includes data collected by the TESS mission. Funding for the TESS mission is provided by the NASA Explorer Program. This paper uses observations made at the South African Astronomical Observatory (SAAO). JA is supported by an STFC studentship. SLC acknowledges support from an STFC Ernest Rutherford Fellowship. RW and PW acknowledge support from STFC (consolidated grant ST/P000495/1). JIV acknowledges support of CONICYT-PFCHA/Doctorado Nacional-21191829.
This project has also received funding from the European Research Council (ERC) under the European Union's Horizon 2020 research and innovation programme (grant agreement No 681601).

%%%%%%%%%%%%%%%%%%%%%%%%%%%%%%%%%%%%%%%%%%%%%%%%%%



\bibliographystyle{mnras}
\bibliography{ref} % if your bibtex file is called example.bib

%%%%%%%%%%%%%%%%%%%%%%%%%%%%%%%%%%%%%%%%%%%%%%%%%%

%%%%%%%%%%%%%%%%% APPENDICES %%%%%%%%%%%%%%%%%%%%%

\appendix

%%%%%%%%%%%%%%%%%%%%%%%%%%%%%%%%%%%%%%%%%%%%%%%%%%

% Don't change these lines
\bsp	% typesetting comment
\label{lastpage}
\end{document}